\documentclass[
    ,final            
    ,numberedheadings 
  ]
  {aipproc}

\layoutstyle{6x9}

\begin{document}

\title{Non-photonic emission from $\gamma$-ray bursts}

\classification{98.70.Rz, 98.70.Sa, 95.85.Ry,    14.60.Pq}
\keywords{gamma ray bursts, cosmic rays, neutrinos}

\author{E. Waxman}{address={Physics Faculty, Weizmann Inst. of Science, Rehovot 76100, Israel}}

\begin{abstract}

$\gamma-$ray bursts (GRBs) are likely sources of ultra-high energy, $>10^{19}$~eV, protons and high energy, $>1$~TeV, neutrinos. Large volume detectors of ultra high energy cosmic rays (UHECRs) and high energy neutrinos, which are already operating and are being expanded, may allow to test in the coming few years the predictions of the GRB model for high energy proton and neutrino production. Detection of the predicted signals will allow to identify the sources of UHECRs and will provide a unique probe, which may allow to resolve open questions related to the underlying physics of GRB models. Moreover, detection of GRB neutrinos will allow to test for neutrino properties (e.g., flavor oscillations for which $\tau\ ^\prime$s would be a unique signature, and coupling to gravity) with an accuracy many orders of magnitude better than is currently possible.

\end{abstract}

\maketitle

\section{Introduction and summary}

The cosmic-ray spectrum extends to energies $\sim10^{20}$~eV, and is likely dominated beyond $\sim10^{19}$~eV by extra-Galactic sources of protons \cite{CR_data_rev} (see, however, \cite{WatsonComposition}). The origin of the highest energy, $>10^{19}$~eV, cosmic rays (UHECRs) is a mystery. As explained in \S\ref{sec:acc}, the stringent constraints, which are imposed on the properties of possible UHECR sources by the high energies observed, rule out almost all source candidates, and suggest that $\gamma$-ray bursts (GRBs) and active galactic nuclei (AGN) are the most plausible sources. In \S~\ref{sec:gzk} we show that the energy loss of UHECR protons due to interaction with microwave background photons, which limits the proton propagation distance, disfavors AGNs as UHECR sources. It is also shown that both the energy production rate and the spectrum of UHECRs are consistent with those expected in the GRB model for UHECR production. Unique predictions of the GRB model, which may be tested with large UHECR detectors, are discussed in \S~\ref{sec:CR_pred} (for a pedagogical review of UHECR production in GRBs see \cite{Wrev}).

High energy neutrino production is discussed in \S~\ref{sec:nu}. It is first shown in \S~\ref{sec:WB} that cosmic ray observations set an upper limit of $E_\nu^2\Phi_\nu\le E_\nu^2\Phi_\nu^{\rm WB}=5\times10^{-8}{\rm GeV/cm^2s\,sr}$ on the diffuse extra-Galactic high energy neutrino intensity produced by sources which, like GRBs and AGN jets, are optically thin for high-energy nucleons to $p\gamma$ and $pp(n)$ interactions. The implications of the upper bound to the detector size required for detecting extra-Galactic high energy neutrinos are briefly discussed. In \S\ref{sec:generic} we show that production of 100~TeV neutrinos in the region where GRB $\gamma$-rays are produced is a generic prediction of the GRB fireball model (for detailed reviews of the fireball model see \cite{fireballs}; for a pedagogical review of high energy neutrino production in GRBs see \cite{Wrev}). 100~TeV neutrino production is a direct consequence of the {\it assumptions} that energy is carried from the underlying engine, most likely a (few) solar mass black hole, as kinetic energy of protons and that $\gamma$-rays are produced by synchrotron emission of shock accelerated electrons. The detection of the predicted neutrino signal will therefore provide strong support for the validity of underlying model assumptions, which is difficult to obtain using photon observations (due to the high optical depth in the vicinity of the GRB "engine"). The predicted neutrino intensity, $\approx0.2\Phi_\nu^{\rm WB}$, implies a detection of $\sim20$ muon induced neutrino events per yr in a km-scale neutrino detector. Since these events should be correlated in time and direction with GRB $\gamma$-rays, the search for GRB neutrinos is essentially background free. Neutrinos may be produced also in other stages of fireball evolution, at energies different than 100~TeV. The production of these neutrinos is dependent on additional model assumptions. As an example, we discuss in \S\ref{sec:astro} the production of TeV neutrinos expected in the "collapsar" scenario, where GRB progenitors are associated with the collapse of massive stars.

The discussion of GRB neutrino emission demonstrates that in addition to identifying the sources of UHECRs, high-energy neutrino telescopes can also provide a unique probe of the physics of these sources. Moreover, detection of high energy neutrinos from GRBs may also provide information on fundamental neutrino properties \cite{WnB97}. High energy neutrinos are expected to be produced in astrophysical sources by the decay of charged pions, which lead to the production of neutrinos with flavor ratio $\Phi_{\nu_e}:\Phi_{\nu_\mu}:\Phi_{\nu_\tau}=1:2:0$ (here $\Phi_{\nu_l}$ stands for the combined flux of $\nu_l$ and $\bar\nu_l$). Neutrino oscillations then lead to an observed flux ratio on Earth of
$\Phi_{\nu_e}:\Phi_{\nu_\mu}:\Phi_{\nu_\tau}=1:1:1$ \cite{Learned:1994wg}. Up-going $\tau$'s, rather than $\mu$'s, would be a distinctive signature of such oscillations, and it has been pointed out
\cite{Learned:1994wg,Athar:2000yw} that searching for deviations from the standard flavor ratio $1:1:1$ may enable one to probe new physics. Detection of neutrinos from GRBs could, moreover, be used to test the simultaneity of neutrino and photon arrival to an accuracy of $\sim1{\rm\ s}$, checking the assumption of special relativity that photons and neutrinos have the same limiting speed. These observations would also test the weak equivalence principle, according to which photons and neutrinos should suffer the same time delay as they pass through a gravitational potential. With $1{\rm\ s}$ accuracy, a burst at $1{\rm\ Gpc}$ would reveal a fractional difference in limiting speed of $10^{-17}$, and a fractional difference in gravitational time delay of order $10^{-6}$ (considering the Galactic potential alone). Previous applications of these ideas to supernova 1987A (see \cite{jnb_book} for review), yielded much weaker upper limits: of order $10^{-8}$ and $10^{-2}$ respectively.

In order to illustrate the possibilities that may be opened by high energy neutrino observations, we show in \S~\ref{sec:nu_phys} that electromagnetic energy losses of $\pi$'s and $\mu$'s modify the flavor ratio (measured at Earth) of neutrinos produced by $\pi$ decay, $\Phi_{\nu_e}:\Phi_{\nu_\mu}:\Phi_{\nu_\tau}$, from $1:1:1$ at low energy to $1:1.8:1.8$ at high energy. For GRBs the transition is expected at $\sim100$~TeV, and may be detected by km-scale $\nu$ telescopes. While the detection of $\tau$ neutrinos would test flavor oscillations, a measurement of the flavor transition energy and energy-width will provide a unique probe of the physical conditions in the $\gamma$-ray production region.

\section{Ultra high energy cosmic rays}
\label{sec:UHECR}

\subsection{The acceleration challenge}
\label{sec:acc}

Most models of particle acceleration in astrophysical sources involve the acceleration of charged particles by the electro-motive force produced through the motion of a magnetized plasma. General phenomenological considerations imply that, independent of the details of the acceleration mechanism, the minimum energy output from a source capable of accelerating in this manner a proton to energy $E_p$ is \cite{Waxman_Nobel}
\begin{equation}\label{eq:L}
  L>\frac{\Gamma^2}{\beta}\left(\frac{E_p}{e}\right)^2c
  =10^{45.5}\frac{\Gamma^2}{\beta}\left(\frac{E_p}{10^{20}\rm eV}\right)^2{\rm erg/s}.
\end{equation}
Here, $u=\beta c$ is the characteristic plasma velocity, and $\Gamma=(1-\beta^2)^{-1/2}$. Only two types of sources are known to satisfy this requirement. The brightest steady sources are active galactic nuclei (AGN). For them $\Gamma$ is typically between 3 and 10, implying $L>10^{47}{\rm erg/s}$, which may be satisfied by the brightest AGN \cite{Lovelace}. The brightest transient sources are GRBs. For these sources $\Gamma\simeq10^{2.5}$ implying $L>10^{50.5}{\rm erg/s}$, which is generally satisfied since the typical observed MeV-photon luminosity of these sources is $L_\gamma\sim10^{52}{\rm erg/s}$. 

The constraint of eq.~(\ref{eq:L}) is essentially obtained by requiring the acceleration time to be shorter than the proton confinement time. A second constraint is imposed by requiring the proton acceleration time to be smaller than its energy loss time. For a relativistic wind, where acceleration takes place in internal shocks arising from variability of the underlying source driving the wind on time scale $\Delta t$, the proton acceleration time is smaller than its synchrotron energy loss time provided \cite{W95a}
\begin{equation}
\Gamma>130 \left(\frac{E_p}{10^{20}\rm eV}\right)^{3/4}\left(\frac{\Delta t}{10\rm ms}\right)^{-1/4}. 
\label{eq:Gmin}
\end{equation}
This constraint is remarkably similar to that inferred from $\gamma$-ray observations on the relativistic dissipative winds which are assumed, within the context of the fireball model (see \cite{fireballs} for reviews), to produce GRBs: $\Gamma>300$ is implied by the $\gamma$-ray spectrum and by the short variability time, $\Delta t\sim10\rm ms$, through the requirement to avoid high pair-production optical depth. This, combined with the constraint of eq.~(\ref{eq:L}), was one of the two main arguments suggested in \cite{W95a} in support of an association between UHECR and GRB sources.

\subsection{The GZK effect}
\label{sec:gzk}

The energy loss of UHECR protons due to interaction with microwave background photons (the "GZK effect") limits the proton propagation distance to $<100$~Mpc for $E_p>10^{20}$~eV \cite{gzk}. This disfavors AGNs as UHECR sources, since no AGN powerful enough to satisfy the constraint of eq.~(\ref{eq:L}) is known to lie within 100~Mpc from Earth. Since most GRBs are known to reside at redshift $z>1$, it may appear that they too are disfavored by the short, 100~Mpc, propagation distance of the protons. This is, however, not the case. While GRB $\gamma$-rays arrive as a burst, slight deflections of protons by an inter-galactic magnetic field would cause large delays in their arrival time (compared to the photon arrival time). The arrival time delay is accompanied by a comparable arrival time spread \cite{W95a}, 
\begin{equation}
\tau(E_p,D)\approx 10^7\left(\frac{E_p}{10^{20}\rm eV}\right)^{-2}\left(\frac{D}{100\rm Mpc}\right)^2
\frac{\lambda B^2}{10^{-8}\rm Mpc\,G}{\rm yr},
\label{eq:delay}
\end{equation}
where $D$ is the source distance, $B$ and $\lambda$ are the field strength and correlation length, and a limit $B\lambda^{1/2}\le10^{-8}{\rm G\ Mpc}^{1/2}$ is set by Faraday rotation measurements \cite{Kronberg94}. The number of GRBs contributing to the flux above energy $E_p$ at any given time is 
\cite{W95a,MnW96} $N_{\rm GRB}(>E_p)\approx(4\pi/ 5) R_{\rm GRB}D_c^3(E_p)\tau[E_p,D_c(E_p)]$, where $R_{\rm GRB}$ is the local ($z=0$) GRB rate density and $D_c(E_p)$ is the proton propagation distance. Using $R_{\rm GRB}(z=0)\approx 0.5\times{\,10^{-9}\rm Mpc^{-3}~yr^{-1}}$ \cite{GPW03}, we find that a large number of GRBs,
\begin{equation}
N_{\rm GRB}(>E_p)\approx10^4 
\left(\frac{E_p}{10^{20}\rm eV}\right)^{-2}\left[\frac{D_c(E_p)}{100\rm Mpc}\right]^5
\frac{\lambda B^2}{10^{-8}\rm Mpc\,G},
\label{eq:N_GRB}
\end{equation}
may contribute to the flux at any given time.

\begin{figure}
\includegraphics[height=.4\textheight]{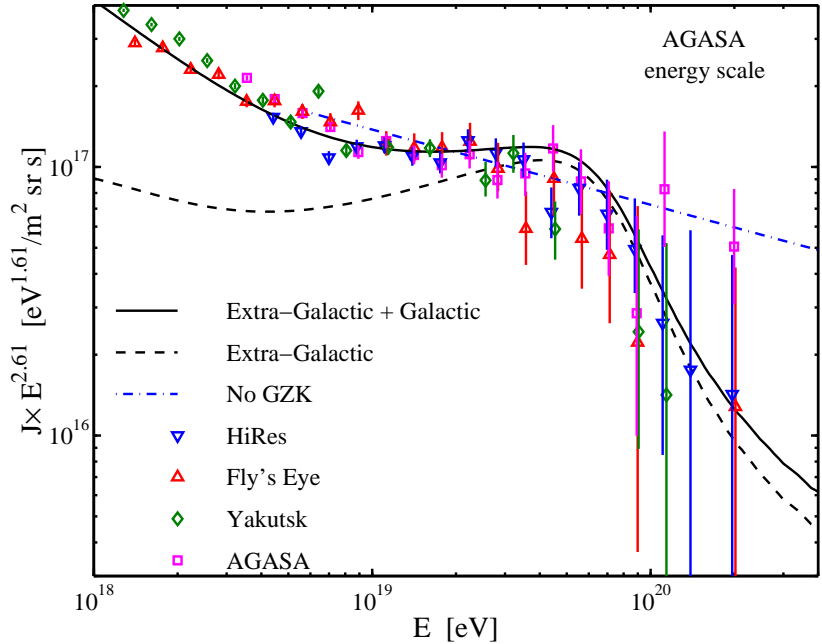}
\caption{The solid curve shows the energy spectrum derived from the two-component model discussed in \S~2.2 (with $\phi(z)\propto (1+z)^3$ up to $z=2$, following the evolution of star formation rate). The dashed curve shows the extra-Galactic component contribution. The "No GZK" curve is an extrapolation of the $E^{-2.75}$ energy spectrum derived for the energy range of $6\times 10^{18}$~eV to $4\times 10^{19}$~eV \cite{CR_data_rev}. Data taken from \cite{fly,exper} (AGASA's energy scale was chosen).}
\label{fig:fig_agasa_norm}
\end{figure}

Figure ~\ref{fig:fig_agasa_norm}, adapted from \cite{BW03}, presents a comparison of available UHECR data 
with the predictions of a model, where extra-galactic protons in the energy range $E_p\le10^{21}$~eV are produced by cosmologically-distributed sources at a rate and spectrum given by
\begin{equation}
E_p^2\frac{d\dot{N}_p}{dE_p}=0.65\times 10^{44} {\rm erg~Mpc^{-3}~yr^{-1}}\phi(z). \label{eq:energyrate}
\end{equation}
Here, $\phi(z)$ accounts for redshift evolution and $\phi(z=0)=1$. The spectrum above $10^{19}$~eV is only weakly dependent on $\phi(z)$ since proton energy loss limits their propagation distance. For the heavy nuclei component dominating at lower, $<10^{19}$~eV, energy the Fly's Eye experimental fit~\cite{fly}, $dN/dE\propto E^{-3.50}$, was used. The power-law spectrum of accelerated particles, $dN/dE\propto E^{-2}$, has been observed for both non-relativistic and relativistic shocks, and is believed to be due to Fermi acceleration in collisionless shocks~\cite{Fermi}. 

Model predictions are in good agreement with the data of all experiments in the energy range $10^{19}$~eV to $10^{20}$~eV (As explained in detail in \cite{BW03}, the various experiments are consistent with each other when systematic errors in the absolute energy scale of the events are taken into account). The suppression of the flux above $\sim10^{19.7}$~eV is the manifestation of the GZK effect. Above $10^{20}$~eV the Fly's Eye, HiRes and Yakutsk experiments are in agreement with each other and with the model, while the AGASA experiment reports a flux higher by a factor $\sim 3$. The origin of this discrepancy is unclear. The Auger UHECR experiment \cite{auger}, which is currently under construction, is expected to significantly reduce the systematic uncertainty in cosmic-ray energy determination and to dramatically increase the number of detected UHECRs (thus reducing statistical errors). It will provide an accurate determination of the UHECR spectrum to energies $>10^{20}$~eV.

The local UHECR energy production rate, eq.~(\ref{eq:energyrate}), is remarkably similar to the local $\gamma$-ray energy production rate by GRBs, $\approx10^{44} {\rm erg~Mpc^{-3}~yr^{-1}}$ \cite{W04}. The similarity of the energy production rates was the second main argument suggested in \cite{W95a} in support of an association between UHECR and GRB sources.

\subsection{GRB model predictions}
\label{sec:CR_pred}

\begin{figure}
\includegraphics[height=.5\textheight]{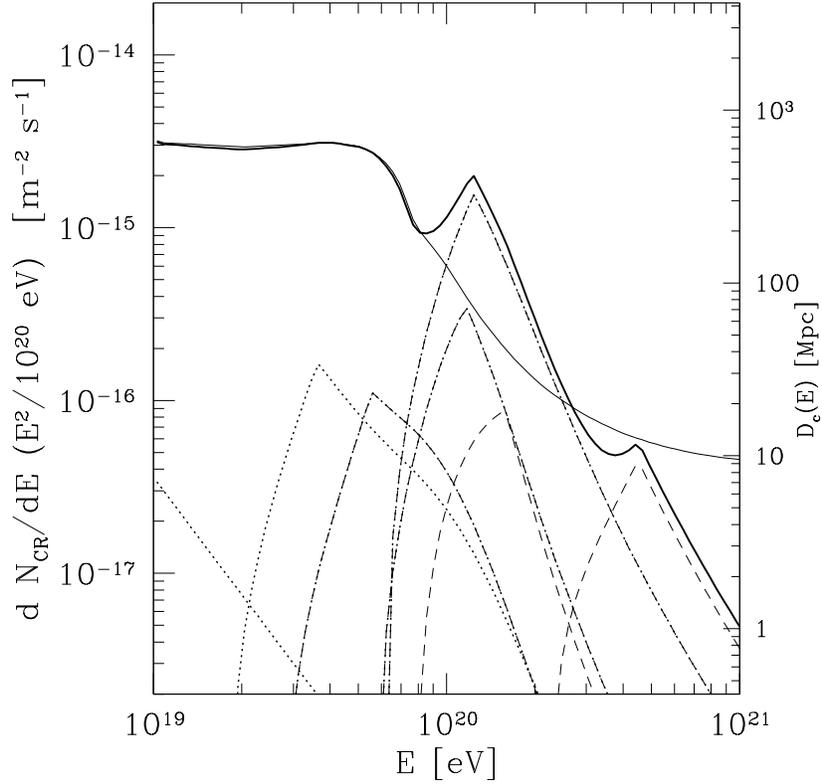}
\caption{Results of a Monte-Carlo realization of the bursting sources
model, with $E_c=1.4\times10^{20}$~eV: Thick solid line- overall 
spectrum in the realization;
Thin solid line- average spectrum, this
curve also gives $D_c(E_p)$;
Dotted lines- spectra of brightest sources at different energies.}
\label{fig:Nc}
\end{figure}
The rapid decrease of $D_c$ with $E_p$ implies, for $B\lambda^{1/2}\le10^{-8}{\rm G\ Mpc}^{1/2}$ and 
$R_{\rm GRB}\le1/{\rm\ Gpc}^3{\rm yr}$, that for some energy $E_c$ in the range $10^{20}{\rm eV}\le E_c<4\times10^{20}{\rm eV}$ the number of GRBs contributing to the UHECR flux is (on average) 1, $N_{\rm GRB}(>E_c)=1$ (see eq.~\ref{eq:N_GRB}). Fig. \ref{fig:Nc} presents the flux obtained for $E_c=1.4\times10^{20}{\rm eV}$ in one realization of a Monte-Carlo simulation described in \cite{MnW96}. For each realization the positions and times at which cosmological GRBs occurred were randomly drawn. Most of the realizations gave an overall spectrum similar to that obtained in the realization of Fig. \ref{fig:Nc} when the brightest source of this 
realization (dominating at $10^{20}{\rm eV}$) is not included. At $E_p < E_c$, the number of sources contributing to the flux is very large, and the UHECR flux received at any given time is near the average (the average flux is that obtained when the UHECR emissivity is spatially uniform and time independent).
At $E_p > E_c$, the flux will generally be much lower than the average, because there will be no burst within a distance $D_c(E_p)$ having taken place sufficiently recently. There is, however, a significant probability to observe one source with a flux higher than the average. A source similar to the brightest one in Fig. \ref{fig:Nc} appears $\sim5\%$ of the time. 

At any fixed time a given burst is observed in UHECRs only over a narrow range of energy, because if
a burst is currently observed at some energy $E_p$, then UHECRs of much lower energy from this burst have not yet arrived,  while higher energy UHECRs reached us mostly in the past. Thus, bursting UHECR sources should have narrowly peaked energy spectra, and the brightest sources should be different at different energies. For steady state sources, on the other hand, the brightest source at high energies should also be the brightest one at low energies. The large exposure expected for the Auger experiment \cite{auger} may allow to distinguish between the two cases, and, if the signature of bursting sources is detected, to determine the bursting source properties (e.g. rate, energy production).

\section{High energy neutrinos}
\label{sec:nu}

\subsection{An upper bound}
\label{sec:WB}

\begin{figure}
\includegraphics[height=.3\textheight]{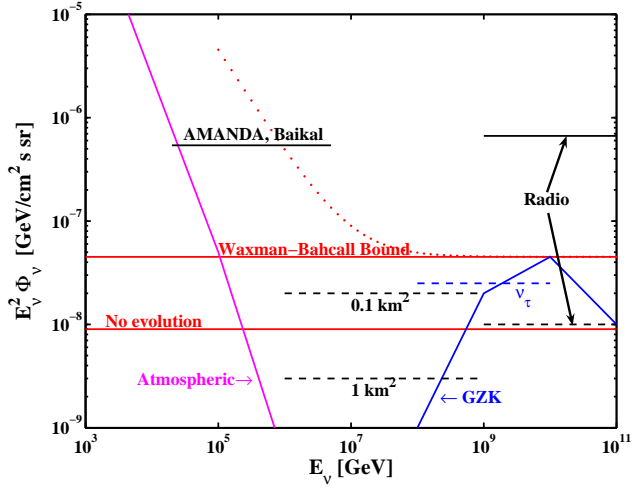}
\caption{The upper bound on the extra-Galactic muon and tau neutrino flux (lower-curve: no source evolution, upper curve: source evolution following star formation rate), assuming oscillations lead to $\Phi_{\nu_e}:\Phi_{\nu_\mu}:\Phi_{\nu_\tau}=1:1:1$, compared with experimental upper bounds (solid lines) of optical Cerenkov experiments (BAIKAL \cite{Baikal}, AMANDA \cite{amanda_bound}), and of coherent Cerenkov radio experiments (RICE \cite{RICE}, GLUE \cite{GLUE}; see also \cite{ANITA}). The curve labelled "GZK" shows the intensity due to interaction with micro-wave background photons. Dashed curves show the expected sensitivity of 0.1~Gton (AMANDA, ANTARES, NESTOR) and 1~Gton (IceCube, NEMO) optical Cerenkov detectors \cite{Nu_telescopes}, of the coherent radio Cerenkov (balloon) experiment ANITA \cite{ANITA} and of the Auger air-shower detector (sensitivity to $\nu_\tau$) \cite{Saltzberg}. Space air-shower detectors (OWL-AIRWATCH) may also achieve the sensitivity required to detect fluxes lower than the WB bound at energies $>10^{18}$~eV \cite{Saltzberg}.}
\label{fig:WBbound}
\end{figure}

The energy production rate, eq.~\ref{eq:energyrate}, sets an upper bound to the neutrino intensity produced by sources which, like GRBs and AGN jets, are optically thin for high-energy nucleons to $p\gamma$ and $pp(n)$ interactions. For sources of this type, the energy generation rate of neutrinos can not exceed the energy generation rate implied by assuming that all the energy injected as high-energy protons is converted to pions (via $p\gamma$ and $pp(n)$ interactions). The resulting upper bound (for muon and anti-muon neutrinos, neglecting mixing) is \cite{WBbound}
\begin{equation}
E_\nu^2\Phi_\nu<2\times10^{-8}\xi_z\left[\frac{(E_p^2d\dot{N}_p/dE_p)_{z=0}}{10^{44}{\rm erg/Mpc^3yr}}\right]{\rm GeV\,cm}^{-2}{\rm s}^{-1}{\rm sr}^{-1}.
\label{eq:WB}
\end{equation}
$\xi_z$ is (a dimensionless parameter) of order unity, which depends on the redshift evolution of $E_p^2d\dot{N}_p/dE_p$ (see eq.~\ref{eq:energyrate}). In order to obtain a conservative upper bound, we adopt $(E_p^2d\dot{N}_p/dE_p)_{z=0}=10^{44}{\rm erg/Mpc^3yr}$ and a rapid redshift evolution, $\Phi(z)=(1+z)^3$ up to $z=2$, following the evolution of star formation rate. This evolution yields $\xi_z\approx3$. The WB upper bound is compared in fig.~\ref{fig:WBbound} with current experimental limits, and with the expected sensitivity of planned neutrino telescopes. 
The figure indicates that km-scale (i.e. giga-ton) neutrino telescopes are needed to detect the expected extra-Galactic flux in the energy range of $\sim1$~TeV to $\sim1$~PeV, and that much larger effective volume is required to detect the flux at higher energy.

\subsection{"Generic" 100~TeV fireball neutrinos}
\label{sec:generic}

Protons accelerated in the fireball internal shocks, where GRB $\gamma$-rays are expected to be produced, lose energy through photo-production of mesons in interactions with fireball photons. The decay of charged pions produced in this interaction results in the production of high energy neutrinos. The key relation is between the observed photon energy, $E_\gamma$, and the accelerated proton's energy, $E_p$, at the threshold of the
$\Delta$-resonance. In the observer frame,
\begin{equation}
E_\gamma \,E_{p} = 0.2 \, {\rm GeV^2} \, \Gamma^2\,.
\label{eq:keyrelation}
\end{equation}
For $\Gamma\approx10^{2.5}$ and $E_\gamma=1$~MeV, we see that
characteristic proton energies $\sim 10^{16}$~eV are required to
produce pions. Since neutrinos produced by pion decay typically
carry $5\%$ of the proton energy, production of $\sim 10^{14}$~eV
neutrinos is expected \cite{WnB97}.

The fraction of energy lost by protons to pions, $f_\pi$, is $f_\pi\approx0.2$ \cite{WnB97,Wrev}. Assuming that GRBs generate the observed UHECRs, the expected
GRB muon and anti-muon neutrino flux may be estimated using eq.~\ref{eq:WB} \cite{WnB97,WBbound}, 
\begin{equation}
E_\nu^2\Phi_{\nu}\approx
0.8\times10^{-8}{f_\pi\over0.2}{\rm GeV\,cm}^{-2}{\rm s}^{-1}{\rm
sr}^{-1}. \label{eq:JGRB}
\end{equation}
This neutrino spectrum extends to $\sim10^{16}$~eV, and is suppressed at higher energy due to energy loss of pions and muons \cite{WnB97,RnM98,WBbound}. Eq.~\ref{eq:JGRB} implies a detection rate of $\sim20$ neutrino-induced muon events per year (over $4\pi$~sr) in a cubic-km detector \cite{WnB97,Alvarez}. Since GRB neutrino events are correlated both in time and in direction with gamma-rays, their detection is practically background free.

\subsection{TeV neutrinos}
\label{sec:astro}

The 100~TeV neutrinos discussed in the previous section are produced at the same region where GRB $\gamma$-rays are produced. Their production is a generic prediction of the fireball model. It is a direct consequence of the assumptions that energy is carried from the underlying engine as kinetic energy of protons and that $\gamma$-rays are produced by synchrotron emission of shock accelerated particles. Neutrinos may be produced also in other stages of fireball evolution, at energies different than 100~TeV. The production of these neutrinos is dependent on additional model assumptions. We discuss below some examples related to the GRB progenitor. For a more detailed discussion see \cite{Mrev,Wrev} and references therein.

The most widely discussed progenitor scenarios for long-duration GRBs involve core collapse of massive stars. In these "collapsar" models, a relativistic jet breaks through the stellar envelope to produce a GRB. For extended or slowly rotating stars, the jet may be unable to break through the envelope. Both penetrating (GRB producing) and "choked" jets can produce a burst of  $\sim5$~TeV neutrinos by interaction of accelerated protons with jet photons, while the jet propagates in the envelope \cite{choked,beacom}. The estimated event rates may exceed $\sim10^2$ events per yr in a km-scale detector, depending on the ratio of non-visible to visible fireballs. A clear detection of non-visible GRBs with neutrinos may be difficult due to the low energy resolution for muon-neutrino events, unless the associated supernova photons are detected. In the two-step "supranova" model, interaction of the GRB blast wave with the supernova shell can lead to detectable neutrino emission, either through nuclear collisions with the dense supernova shell or through interaction with the intense supernova and backscattered radiation field \cite{supranova}.

\subsection{Energy dependent neutrino flavor ratio}
\label{sec:nu_phys}

\begin{figure}
\includegraphics[height=.25\textheight]{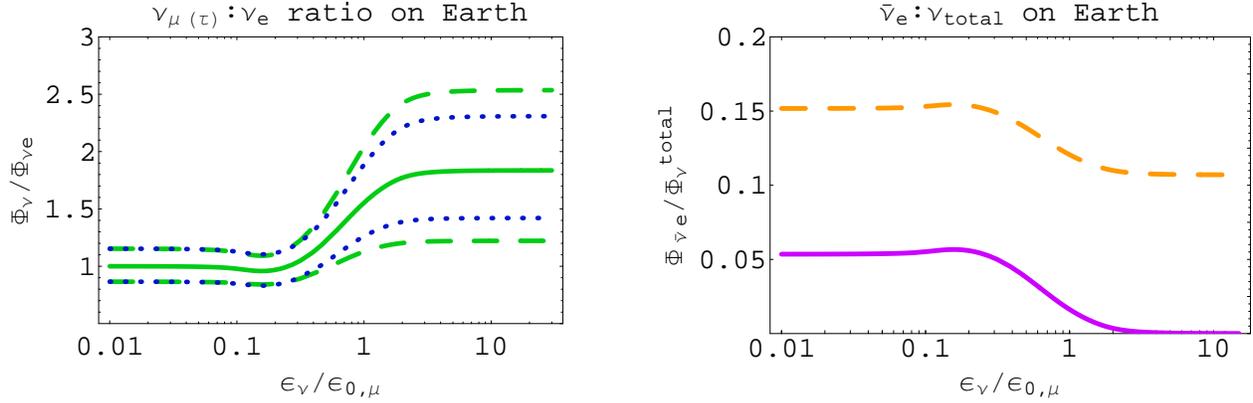}
\caption{Flavor and anti-particle content of the flux of astrophysical neutrinos produced by pion decay, for pion energy spectrum (at production) $dn/dE_\pi\propto E_\pi^{-2}$ and electromagnetic pion energy loss rate $d E_\pi/dt\propto E_\pi^2$, as expected for internal shocks in GRB fireballs. Left: The ratio between $\Phi_{\nu_{\mu}(\nu_\tau)}$ and $\Phi_{\nu_e}$ (solid line), with 90\% CL lines of $\nu_\mu$ (dashed) and $\nu_\tau$ (dotted) fluxes (here, $\Phi_{\nu_l}$ stands for the combined flux of $\nu_l$ and $\bar\nu_l$). Right: The ratio of $\bar\nu_e$ to total $\nu$ flux on Earth, solid (dashed) line for neutrinos produced by  $p\gamma$ ($pp$) interactions. $E_{0,\mu}$ is the muon energy for which the muon life time is comparable to its electromagnetic energy loss time. $E_{0,\mu}\approx10^3$~TeV for internal shocks in GRB fireballs.}
\label{fig:fNu}
\end{figure}

Although a $1:1:1$ flavor ratio appears to be a robust prediction of models where neutrinos are produced by pion decay, energy dependence of the flavor ratio is a generic feature of models of high energy astrophysical neutrino sources \cite{Kashti}. Pions are typically produced in environments where they may suffer significant energy losses prior to decay, due to interaction with radiation and magnetic fields \cite{WnB97,RnM98}. Since the pion life time is shorter than the muon's, at sufficiently high energy the probability for pion decay prior to significant energy loss is higher than the corresponding probability for muon decay. This leads to suppression at high energy of the relative contribution of muon decay to the neutrino flux. The flavor ratio is modified to $0:1:0$ at the source, implying $1:1.8:1.8$ ratio on
Earth \cite{Kashti}.

Figure~\ref{fig:fNu} presents the expected energy dependence of flavor and anti-particle content for internal shocks in GRB fireballs. Since the transition is expected at $\sim100$~TeV, it may be
detected by km-scale $\nu$ telescopes.

\end{document}